\documentclass[lettersize,journal]{IEEEtran}
\usepackage{url}
\usepackage{amsfonts}       % blackboard math symbols
\usepackage{nicefrac}       % compact symbols for 1/2, etc.
\usepackage{mathtools}      % microtypography
\usepackage{color,bm}
\usepackage{mathrsfs}%huati font
\usepackage{amsmath}%split
\usepackage[linesnumbered,ruled,vlined]{algorithm2e}%algorithm
\usepackage{graphicx} 
\usepackage{multirow}
\usepackage{amssymb}
\usepackage{mathtools}
\usepackage{amsthm}
\usepackage[skins]{tcolorbox}

\newtheorem{definition}{Definition}
\newtheorem{theorem}{Theorem}
\newtheorem{lemma}{Lemma}
\newtheorem{proposition}{Proposition}
\newtheorem{corollary}{Corollary}

\title{Inverted Inference and Recursive Bootstrapping: \\A Primal–Dual Theory of Structured Cognition}

\author{%
  Xin Li,\textsuperscript{1}\thanks{This work was partially supported by NSF IIS-2401748 and BCS-2401398. The author has used ChatGPT models to assist in the development of theoretical ideas presented in this paper.} \\ Department of Computer Science\\
  University at Albany\\
  \texttt{xli48@albany.edu}
}
\begin{document}

\maketitle

\begin{abstract}

This paper presents a unifying framework that links cognitive representation under the Context--Content Uncertainty Principle (CCUP) with optimal transport (OT) through the lens of primal-dual inference. We propose that representations function not merely as data structures, but as active dual constraints that define feasible manifolds over which learning and inference operate. Cognition is thus formalized as a process of dynamic alignment between high-entropy contexts and low-entropy content, achieved through cycle-consistent inference that minimizes conditional entropy.
Building on this formulation, we introduce \textit{inverted inference} as the core mechanism for goal-directed simulation. Inspired by the structure-before-specificity principle, inverted inference reverses the direction of conditioning to simulate latent trajectories consistent with internal goals. This asymmetric cycle closes the duality gap in constrained optimization by aligning context (primal variables) with content (dual constraints), thereby transforming inference into structure-constrained entropy minimization.
We extend this framework temporally through \textit{recursive bootstrapping}, where each inference cycle refines the constraint manifold for the next, enabling path-dependent optimization via memory chains. This temporal organization supports hierarchical goal decomposition and planning through structured subgoal simulation. Spatially, we generalize the model via \textit{hierarchical spatial bootstrapping}, connecting it to the Hierarchical Navigable Small World (HNSW) graph structure. This allows fast, sublinear retrieval of goal-consistent states by navigating structured latent manifolds.
Together, these contributions yield a computationally grounded theory of cognition in which dynamic alignment, across time and space, enables efficient generalization, abstraction, and goal-directed behavior. We conclude that CCUP provides a scalable architecture for both slow, recursive reasoning and fast, structure-aware recognition through layered primal-dual optimization.

\end{abstract}


\section{Introduction}
\label{sec:1}

Classical models of neural computation often assume that the brain optimizes cost functions or updates representations via gradient flow in a parameter space. However, these models face fundamental limitations in scalability, tractability, and biological plausibility \cite{bengio2013representation}. By contrast, neural systems consistently extract structure, infer latent causes, and generalize across novel situations, despite operating under severe biological constraints (e.g., noisy and partial observability). This success is not due to raw computational power (e.g., mammalian brains have limited power), but rather to a set of organizing principles that govern how representations are formed, constrained, and reused \cite{mead2012analog}. 

We propose that one such foundational principle is the \textbf{Context–Content Uncertainty Principle} (CCUP) \cite{li2025CCUP}, which frames cognition as a cyclic process of uncertainty minimization. Under CCUP, the brain resolves the inherent ambiguity of high-entropic inputs (context) by dynamically binding them to structured, low-entropic representations (content) and uses these representations to guide future perception and action \cite{fuster2004upper}. This bidirectional inference mechanism offers a unified explanation for how the brain achieves efficient learning and planning by operating on constrained manifolds rather than exhaustively searching through raw state spaces \cite{friston2017active}.

While CCUP characterizes cognition as a cycle that aligns high-entropy context with low-entropy content through bidirectional inference, it also raises a deeper question: how is such content made available in the first place? Efficient inference presupposes the existence of structured representational manifolds, spaces where possible interpretations are already constrained before specific sensory inputs arrive. This motivates the \textbf{structure-before-specificity} principle \cite{clark2013whatever}: the brain does not build structure from data alone; rather, it first constructs abstract, low-dimensional scaffolds (e.g., schemas and goal-relevant latent spaces \cite{niv2019learning}) onto which incoming observations are subsequently mapped. In this view, structure precedes data fitting, allowing the system to avoid the curse of dimensionality \cite{bellman1966dynamic} by narrowing the search space to a task-relevant, semantically organized subset. This principle operationalizes CCUP by explaining how content can remain low entropy even before sensory disambiguation: it is seeded by structure.

When cast under an optimization framework, the brain optimizes \emph{over representations} in the latent space, and representations themselves are \emph{constraints} abstracted from data. Such observation inspires us to develop a theoretical framework that integrates the CCUP with the mathematical theory of constrained optimization, drawing particularly from optimal transport \cite{villani2009optimal} and the primal-dual method \cite{wright1997primal}. This leads to two new insights about the interplay between representation and optimization: 1) \textbf{representations as constraint manifolds.} Rather than encoding data points, representations define \emph{low-dimensional manifolds} within a high-dimensional ambient space. These manifolds constrain the domain of inference and learning, enabling generalization by ruling out irrelevant degrees of freedom \cite{tenenbaum2000global}. 2) \textbf{optimization via inverted inference.} In contrast to forward inference from observation to latent cause, \emph{inverted inference} proceeds from internally specified goals or priors to constrain possible interpretations or actions. This process acts as a \emph{dual constraint} on inference, functionally analogous to dual variables in optimization theory \cite{friston2009reinforcement}.

Together, these principles offer a computationally grounded and geometrically informed view of neural intelligence. By treating representations as active, structured constraints rather than passive encodings, we reframe cognition as constrained optimization on goal-conditioned manifolds. Temporal bootstrapping through recursive memory chains supports \emph{slow thinking} \cite{kahneman2011thinking}, a deliberative mode of cognition that unfolds through path-dependent inference and structure refinement over time. In parallel, spatial bootstrapping enables \emph{fast thinking} \cite{kahneman2011thinking} by leveraging hierarchical priors to rapidly align sensory input with top-down expectations, enabling sublinear recognition and generalization. This unified framework integrates key ideas from information geometry \cite{amari2016information}, variational inference \cite{kingma2014auto}, and neural computation \cite{hertz2018introduction}, providing a common theoretical umbrella for understanding the dual modes of cognition as layered primal--dual inference processes operating across time and space.
The contributions of this paper are summarized below.
\begin{enumerate}
    
    \item \textbf{Dynamic alignment between context and content representations.} Under CCUP, cognition is formulated as a dynamic alignment process that minimizes the conditional entropies between context and content. Each cycle alternates between bottom-up disambiguation and top-down reconstruction. Aligning internal representations with external data through convergence on a constraint manifold is shown to eliminate the information bottleneck for optimal transport.

    \item \textbf{Inverted inference closes the duality gap.} Inspired by the structure-before-specificity principle, we construct an asymmetric cycle with inverted inference for goal-directed simulation, which generalizes the half-step down trick in Q-learning. This inverted direction acts as a dual constraint, structurally analogous to the dual variables in constrained optimization. We show that the \emph{duality gap} in primal-dual optimization corresponds to the divergence between context and content. A vanishing duality gap indicates that context (primal input) is fully constrained by and consistent with its representation (dual constraint).

    \item \textbf{Path-dependent optimization via recursive temporal bootstrapping.} Unlike static manifold learning, the CCUP framework supports \emph{path-dependent inference}, where each inference cycle updates the constraint manifold that governs the next. This recursive temporal bootstrapping extends the asymmetric cycle into the memory chain, linking cognition to planning and imagination. Path-dependent inference leads to a dynamic theory of slow thinking where convergence arises from cycle completion, not just from gradient flow.

    \item \textbf{Small-world optimal transport via hierarchical spatial bootstrapping.} We also extend inverted inference spatially by connecting it with a hierarchical navigable small network (HNSW) model. A hierarchical extension of dynamic alignment enables goal-directed simulation for spatial inference, which is equivalent to layered primal-dual optimization. HNSW operationalizes spatial inverted inference efficiently by embedding the manifold $\mathcal{M}_{\text{goal}}$ in a navigable topological graph, allowing goal-conditioned subspace retrieval with sublinear cost.

\end{enumerate}

\section{Background: an Information Geometric Framework for Representation and Optimization}
\label{sec:1}

At the heart of cognition lies the task of interpreting and acting upon ambiguous sensory and internal signals under uncertainty \cite{knill2004bayesian}. A central challenge is how to form stable, goal-relevant representations (\emph{content}) from high-entropic, variable input (\emph{context}). 
The \textbf{Context-Content Uncertainty Principle (CCUP)} formalizes this challenge by characterizing cognition as a dynamic alignment between \emph{context} (high-entropy, ambiguous conditions, denoted \( \Psi \)) and \emph{content} (low-entropy, specific representations, denoted \( \Phi \)) \cite{li2025CCUP}. Formally, we start with the following definition.

\begin{definition}[Context–Content Alignment]
Let \( \Psi \) and \( \Phi \) be random variables representing context and content, respectively, with associated inference map \( q(\Phi | \Psi) \) and generative map \( p(\Psi | \Phi) \). We say that the pair \( (\Psi, \Phi) \) is \emph{aligned} if the following cycle-consistency conditions hold for some error tolerance \( \varepsilon \geq 0 \):
\begin{align*}
D_{\mathrm{KL}}\left( P(\Psi) \,\Vert\, \tilde{P}_\Psi \right) &\leq \varepsilon,  \tilde{P}_\Psi := \mathbb{E}_{\Psi \sim P(\Psi)} \left[ p(\Psi | \hat{\Phi}(\Psi)) \right], \\
D_{\mathrm{KL}}\left( P(\Phi) \,\Vert\, \tilde{P}_\Phi \right) &\leq \varepsilon,  \tilde{P}_\Phi := \mathbb{E}_{\Phi \sim P(\Phi)} \left[ q(\Phi | \hat{\Psi}(\Phi)) \right],
\end{align*}
where \( \hat{\Phi}(\Psi) \sim q(\Phi | \Psi) \) and \( \hat{\Psi}(\Phi) \sim p(\Psi | \Phi) \) are samples drawn from the respective mappings.
Additionally, alignment requires high mutual information between context and content:
$I(\Psi; \Phi) \approx H(\Psi) \approx H(\Phi)$,
ensuring that the inference and generative processes capture nearly all relevant structure.
\end{definition}

\noindent \textbf{Remark:} Context-content alignment defines a cycle-consistent inference structure in which context and content mutually constrain one another with minimal informational loss.
Importantly, this alignment is \emph{not symmetric} \cite{anderson1972more}: context often occupies a compressed sensory or environmental space but remains ambiguous due to aliasing and interference, whereas content spans a semantically rich latent space but is shaped by goal-driven constraints that stabilize its entropy \cite{spreng2010default}.
CCUP asserts a fundamental trade-off in inference: the system seeks to minimize both 1) \textbf{specification}: conditional entropies \( H(\Phi | \Psi) \) (context disambiguation); and 2) \textbf{generalization}: \( H(\Psi | \Phi) \)  (context reconstruction). The main result of CCUP is stated as follows (its proof can be found in Appendix A):

\begin{lemma}[CCUP Lower Bound]
Let \( \Psi \) and \( \Phi \) be random variables representing context and content. Then, we have
$H(\Phi | \Psi) + H(\Psi | \Phi) \geq H(\Psi, \Phi) - I(\Psi; \Phi)$
with equality if and only if the joint distribution \( p(\Psi, \Phi) \) is cycle-consistent and bidirectionally minimal (i.e., $(\Psi, \Phi)$ are \emph{aligned}).
\end{lemma}

\noindent\textbf{Remark:} The above inequality highlights the irreducible uncertainty in attempting to map between context and content unless a consistent inference cycle is established.
The CCUP lower bound provides a formal backbone for understanding why cognition must \emph{break symmetry}. In real-world conditions, the bidirectional mappings between \( \Psi \) and \( \Phi \) are lossy: perception suffers from aliasing \cite{wehner1987matched}, memory from interference \cite{mccloskey1989catastrophic}, and action from feedback ambiguity \cite{todorov2002optimal}. These conditions make it impossible to achieve a perfect mutual mapping, thereby necessitating structural solutions such as \emph{asymmetric inference cycles} to iteratively minimize uncertainty, which we call ``dynamic alignment'' in this paper.

\noindent\textbf{Information geometric perspective.} In both cognitive systems and optimization theory, tractable computation arises not from brute-force search over high-dimensional spaces, but from strategic restriction to low-dimensional, structure-preserving manifolds. We propose that representations, under the CCUP framework, correspond to such manifolds, informationally compressed, geometrically constrained subspaces in which inference is both efficient and generalizable.
This information geometric perspective \cite{amari2016information} clarifies why the brain does not optimize over raw sensory inputs. Instead, it operates on structured, representational manifolds where inference is confined to semantically meaningful and computationally efficient subspaces. Representations, viewed as geometric constraints, thus serve both epistemic and algorithmic roles: they make the world knowable and inference doable.

\begin{lemma}[Representations as Dual Constraints]
Let \( \Psi \) be a high-entropy contextual variable and \( \Phi \) be a structured, low-entropy representation. Let \( \mathcal{M}_\Phi \subset \mathbb{R}^n \) denote the manifold of latent states consistent with \( \Phi \). Then inference under CCUP:
$\min_{p(\Psi, \Phi)} H(\Psi | \Phi) + H(\Phi)$
is equivalent to solving a constrained optimization problem:
$\min_{x} f(x)  ~\text{such that}~  g(x) = \Phi$,
where \( g(x) \) is the representational constraint function, and \( \Phi \) serves as the dual variable shaping feasible inference.
\end{lemma}

\noindent\textbf{Remark:} Structured content representations are equivalent to dual constraints in optimization. Both stabilize uncertainty and convert intractable search into tractable solution manifolds \cite{parr2022active}.
Formally, let the ambient space $\mathcal{X}$ represent all possible configurations of observed data or cognitive states, typically high-dimensional and intractable. A representation $\mathcal{M} \subset \mathcal{X}$ is a submanifold defined implicitly by a set of constraints (e.g., features selected and abstracted through prior experience or learned structure). In the optimization setting, this corresponds to restricting the feasible set of solutions to lie on $\mathcal{M}$, enforced via dual variables associated with each constraint.
The optimization process then becomes a constrained inference problem:
\begin{equation}
\min_{x \in \mathcal{X}} f(x)  ~\text{subject to}~  x \in \mathcal{M},
\end{equation}
which can be reformulated using the primal-dual method as:
\begin{equation}
\min_{x \in \mathcal{X}} \max_{\lambda} \mathcal{L}(x, \lambda) = f(x) + \lambda^\top \phi(x),
\end{equation}
where $\phi(x) = 0$ defines the representational manifold $\mathcal{M}$, and $\lambda$ are the dual variables encoding the context-specific constraints.
In this framework, the dual variables are not merely Lagrange multipliers; they function as contextual forces that align the current state (primal variable) with its representation manifold. The \emph{duality gap} \cite{boyd2004convex},
\begin{equation}
\text{Gap} = \min_{x} \max_{\lambda} \mathcal{L}(x, \lambda) - \max_{\lambda} \min_{x} \mathcal{L}(x, \lambda),
\end{equation}
quantifies the degree of misalignment between context (primal input) and representation (dual constraint). A vanishing duality gap indicates cycle-consistent inference, where the context is fully constrained by, and consistent with, its representation. It follows that Lemma 2 can be extended into the following.

\begin{lemma}[CCUP as an Entropic Duality Principle]
Let \( \Psi \) denote contextual variables (high-entropy) and \( \Phi \) denote content variables (low-entropy). The inference objective under CCUP is to minimize joint uncertainty:
$\min_{p(\Psi, \Phi)} \; H(\Psi | \Phi) + H(\Phi)$
subject to structural constraints imposed by content:
$\mathbb{E}_{p(\Psi, \Phi)}[g(\Psi, \Phi)] = 0$
Then the problem admits a Lagrangian dual form:
$\mathcal{L}(p, \lambda) = H(\Psi | \Phi) + H(\Phi) + \lambda^\top \mathbb{E}_{p}[g(\Psi, \Phi)]$
whose saddle point characterizes optimal cyclical coordination between context and content.
\end{lemma}

\noindent\textbf{Remark:} The CCUP encodes a duality between the variability of contextual inference and the structure of content constraints, revealing that cognition performs joint entropy minimization via dual coordination—analogous to constrained variational optimization. Such information geometric perspective inspires us to connect dynamic alignment with the theory of optimal transport next.

\begin{comment}
\begin{tcolorbox}[title=Interpretation of CCUP as Entropic Duality, colback=blue!5!white, colframe=blue!75!black, fonttitle=\bfseries]
%\textbf{Summary:}
In the geometry of information spaces, content (\( \Phi \)) and context (\( \Psi \)) play dual roles:
1) context corresponds to expectation parameters \( \eta \) (variability across samples).
2) content corresponds to natural parameters \( \theta \) (goal constraints).
%\end{itemize}
Under CCUP, inference minimizes joint entropy:
$H(\Psi ,\Phi)$,
which geometrically corresponds to:
1) projecting a high-entropy context onto a goal-constrained content submanifold.
2) solving for the saddle point of the dual entropy surface, where inference gradients align.
%\end{enumerate}
%This mirrors Lagrangian duality and Legendre transformation in information geometry, revealing that CCUP performs structured variational inference through dual coordination across a statistical manifold. 
%\textbf{Implications:} Representations are not just data; they are constraints. Optimization solves for solutions within these constraint manifolds.
\end{tcolorbox}
\end{comment}

\section{Dynamic Alignment and Optimal Transport}

The inherent asymmetry in entropy between context and content variables makes information loss during one-way transport unavoidable, creating an information bottleneck (IB) \cite{tishby2000information}. The only way to overcome this bottleneck is through dynamic alignment: an inference cycle that co-evolves content and context representations via top-down and bottom-up passes. We first give equivalent conditions for dynamic alignment below.

\begin{definition}[Equivalent Conditions for Context-Content Alignment]
Let $\Psi$ denote a high-entropy context variable and $\Phi$ a low-entropy content variable, with bidirectional inference maps $q(\Phi | \Psi)$ (inference/disambiguation) and $p(\Psi | \Phi)$ (generative/reconstruction). The pair $(\Psi, \Phi)$ is said to be \emph{aligned} if the following equivalent conditions hold:

\begin{enumerate}
    \item \textbf{Cycle Consistency:}
    $D_{\mathrm{KL}}(\Psi \, \Vert \, \hat{\Psi}) \leq \varepsilon,  D_{\mathrm{KL}}(\Phi \, \Vert \, \hat{\Phi}) \leq \varepsilon$,
    where $\hat{\Phi} = q(\Phi | \Psi)$ and $\hat{\Psi} = p(\Psi | \hat{\Phi})$, and $\varepsilon \geq 0$ is a small alignment tolerance.

    \item \textbf{Mutual Information Saturation:}
    $I(\Psi; \Phi) \approx H(\Phi) \approx H(\Psi)$,
    indicating that context and content share nearly all their informational degrees of freedom.

    \item \textbf{Entropy Symmetry (CCUP Condition):}
    $H(\Phi | \Psi) \approx 0,  H(\Psi | \Phi) \approx 0$,
    such that each variable fully determines the other under minimal uncertainty.

    \item \textbf{Optimal Transport Coupling:}
    $\gamma^* = \arg\min_{\gamma \in \Pi(P_\Psi, P_\Phi)} \mathbb{E}_{\gamma}[c(\Psi, \Phi)] - \varepsilon H(\gamma)$,
    where $\gamma^*$ is a near-deterministic transport plan with low entropy, implying functional invertibility between context and content.

    \item \textbf{Duality Gap Closure:}
    $\min_{\Phi} \max_{\lambda} \mathcal{L}(\Phi, \lambda) = \max_{\lambda} \min_{\Phi} \mathcal{L}(\Phi, \lambda)$,
    where $\mathcal{L}(\Phi, \lambda)$ is the Lagrangian associated with constrained optimization, and the zero duality gap indicates consistency between representation and constraint.
\end{enumerate}
\end{definition}

In classical formulations, optimal transport (OT) \cite{villani2009optimal} faces the challenge of mapping a high-entropy contextual distribution \( \mu(\Psi) \) to a lower-entropy content distribution \( \nu(\Phi) \) via a latent variable \( Z \). Traditional wisdom frames this as a one-shot encoding problem: how to encode \( \Psi \) into \( Z \) while preserving relevance to \( \Phi \), which introduces an IB due to the entropy gap \( H(\Psi) \gg H(\Phi) \).
CCUP offers a fundamentally different perspective on introducing a feedback loop to form a dynamic cycle based on the following two observations. 

First, it reframes the problem as \emph{dynamic alignment} between two asymmetric representations. Rather than discarding information, inference under CCUP involves disambiguating \( \Psi \) through bottom-up encoding/conditioning \( p(\Phi | \Psi) \), and reconstructing \( \Psi \) from \( \Phi \) via top-down decoding/reconstruction \( p(\Psi | \Phi) \). These are not opposing or dialectical constraints but mutually supportive operations that iteratively align bilateral (content and context) representations through bidirectional inference \cite{parr2017uncertainty}. 
Second and more crucially, CCUP resolves the bottleneck through \emph{cycle formation}, which transforms the directional arrow of inference into a convergent loop that distributes inference across time, as defined below. 

\begin{definition}[Cycle Formation under CCUP]
With \( \Phi \) and \( \Psi \) as defined before, nature completes the cycle
$\Psi \xrightarrow{\text{encoding}} \Phi \xrightarrow{\text{decoding}} \hat{\Psi} \xrightarrow{\text{update}} \Phi'$
such that:
%\begin{enumerate}
1) {context disambiguation(encoding):} \( \Phi \) is an encoded representation of \( \Psi \) satisfying \( H(\Phi) \ll H(\Psi) \); 2) {context reconstruction(decoding):} A generative process \( G \) expands \( \Phi \) into a prediction \( \hat{\Psi} = G(\Phi) \), leading to prediction error: \( \delta = d(\Psi, \hat{\Psi}) \); and 3) {content update:} the representation is updated as \( \Phi' = \Phi - \eta \nabla_\Phi d(\Psi, \hat{\Psi}) \), reducing uncertainty via gradient descent.
%\end{enumerate}
If this cycle is iterated with bounded reconstruction error \( d(\Psi, \hat{\Psi}) \leq \epsilon \), the process converges to a fixed point where content and context are dynamically aligned, and mutual information \( I(\Phi; \Psi) \) is maximized.
\end{definition}

%Classical formulations of the IB in OT aim to map a high-entropy contextual distribution \( \mu(\Psi) \) to a lower-entropy content distribution \( \nu(\Phi) \) via a latent variable \( Z \), while minimizing transport cost and preserving relevance. This is traditionally viewed as a compression-relevance trade-off, with the encoding \( p(Z |\Psi) \) constrained to retain as much information about \( \Phi \) as possible.However, CCUP reveals that this bottleneck emerges not from compression per se, but from an underlying \emph{asymmetry of directional information loss}. %Specifically, the context-to-content mapping \( p(\Phi I \Psi) \) inherently discards fine-grained contextual detail due to aliasing, while the reverse mapping \( p(\Psi I \Phi) \) requires reintegration over uncertainty to reconstruct the original input.  

The bottom-up encoding and top-down decoding processes must be coordinated through \emph{iterative bidirectional inference} \cite{dayan1995helmholtz,hinton1995wake}, such that the forward and backward mappings form a cycle that aligns representations across inference passes. The cyclic system turns the entropy asymmetry between context and content into a temporal resource, bootstrapping stability through dynamic consistency \cite{friston2017graphical}. Formally, we have

\begin{theorem}[Cycle Consistency Eliminates the IB in OT]
Let \( \Phi \) and \( \Psi \) be as defined before along with associated conditional distributions \( p(\Phi | \Psi) \) and \( p(\Psi | \Phi) \). Suppose that inference proceeds via an iterative cycle:
$\Psi \rightarrow \Phi \rightarrow \Psi$,
where each pass updates representations to reduce uncertainty. If this bidirectional inference process converges to a fixed point such that the forward and backward transport maps become cycle-consistent, i.e., $\Psi \approx \mathbb{E}_{p(\Psi | \Phi)}[\Psi], \Phi \approx \mathbb{E}_{p(\Phi | \Psi)}[\Phi],$
then we have: $H(\Phi | \Psi) \to 0, H(\Psi | \Phi) \to 0,$ and mutual information is fully recovered: $I(\Psi; \Phi) = H(\Psi) = H(\Phi).$
%Thus, the IB is eliminated and OT is achieved through dynamic alignment rather than one-time encoding.
\end{theorem}

\noindent\textbf{Remark:} The proof is referred to Appendix B. This theorem demonstrates that IB in OT is not an inherent limitation of representation capacity but a consequence of misalignment between asymmetric variables. Under CCUP, this misalignment can be overcome through cycle formation, which dynamically aligns representations through bottom-up disambiguation and top-down reconstruction \cite{rao1999predictive}. However, any cycle completion involves a chicken-and-egg problem; to resolve this issue, we state the following fundamental result of broken symmetry under the framework of CCUP \cite{li2025Broken}. 

\begin{theorem}[Structure Precedes Specificity]
With \( \Phi \) and \( \Psi \) as defined before, let \( \mathcal{D} = \{ \Psi_i \}_{i=1}^N \) be a set of observed contexts drawn from an environment. An optimal inference strategy proceeds in two stages:
1) {Generalization}:
    The agent seeks a latent structure \( \Phi \) that minimizes the expected conditional entropy across contexts:
    $\Phi^\ast = \arg\min_{\Phi} \mathbb{E}_{i} \left[ H(\Psi_i | \Phi) \right]$, which yields a generalizable representation \( \Phi^\ast \) that captures invariant structure across variable observations;
2) {Specification}:
    Once \( \Phi^\ast \) is established, the agent performs context-specific inference by minimizing:
    $H(\Phi^\ast | \Psi_j)$ for each new observation \( \Psi_j \in \mathcal{D}_{\text{new}} \), effectively binding content to context using the learned structural prior.
    \label{thm:structure_precedes_specificity}
\end{theorem}

\noindent\textbf{Remark:} The proof is referred to Appendix C. This principle implies that inference does not begin with data-driven search, but with the imposition of low-dimensional, goal-conditioned constraints that shape the space of possible solutions \cite{ma2006bayesian}. This anticipatory structuring naturally gives rise to an asymmetric cycle in OT, where high-entropy context is mapped onto a pre-structured content manifold \cite{peyre2019computational}. Meanwhile, the feedback from reconstruction refines the constraint landscape, forming a directed loop that resolves representational and transport asymmetries. Formally, we have

\begin{theorem}[Structure-Before-Specificity as Asymmetric Cycle Completion in OT]
Let $P_\Psi$ be a high-entropy context distribution over a space $\mathcal{C}$, and let $\mathcal{M}_\lambda \subset \mathcal{R}$ be a goal-conditioned, low-dimensional manifold of content representations parameterized by dual variables $\lambda$. Suppose the agent performs inference via an asymmetric transport cycle consisting of:
1) {structure seeding:} a prior $\lambda$ imposes representational constraints defining $\mathcal{M}_\lambda$; 2) {forward transport:} context $\Psi \sim P_\Psi$ is transported to content $\Phi \in \mathcal{M}_\lambda$ via a coupling $\gamma_\lambda \in \Pi(P_\Psi, P_\Phi^\lambda)$; 3) {backward reconstruction:} the transported $\Phi$ induces a reconstruction $\hat{\Psi}$ via a generative map $p(\Psi | \Phi)$.
Then the following are equivalent:
\begin{enumerate}
    \item The inference cycle is self-consistent:
    $D_{\mathrm{KL}}\left(P_\Psi \, \Vert \, \hat{P}_\Psi \right) \leq \varepsilon,  \text{where}  
\hat{P}_\Psi(\psi) := \int p(\psi | \phi) q(\phi | \psi') P_\Psi(\psi') \, d\psi' \, d\phi$.
       
    \item The duality gap in the Lagrangian
    $\mathcal{L}(\Phi, \lambda) = \mathbb{E}_{P_\Psi}[c(\Psi, \Phi)] + \lambda^\top \phi(\Phi)$
    vanishes:
    $\min_{\Phi} \max_{\lambda} \mathcal{L}(\Phi, \lambda) = \max_{\lambda} \min_{\Phi} \mathcal{L}(\Phi, \lambda)$.
    
    \item The coupling $\gamma_\lambda^*$ solves the entropy-regularized OT problem:
    $\gamma_\lambda^* = \arg\min_{\gamma \in \Pi(P_\Psi, P_\Phi^\lambda)} \mathbb{E}_{\gamma}[c(\Psi, \Phi)] - \varepsilon H(\gamma)$, 
    and satisfies the following constraint:
    $\Phi \in \mathcal{M}_\lambda,  \text{with}  \dim(\mathcal{M}_\lambda) \ll \dim(\mathcal{C})$.
\end{enumerate}

\end{theorem}

\noindent\textbf{Remark:} the definition of $\hat{P}_\Psi(\psi)$ in self-consistent inference cycle is the same as the pushforward distribution in Definition 1.
Under these conditions, the asymmetric cycle converges to a consistent alignment of context and content, and inference becomes tractable through constrained, goal-seeded optimal transport. Next, we study computational aspects of this asymmetric cycle in detail.

\section{Asymmetric Cycle and Inverted Inference}

\subsection{Asymmetric Cycle Breaks the Curse of Dimensionality}

At the core of the computational bottleneck lies the classical search problem: given a vast space of possible actions and states, how can an agent efficiently identify a trajectory that achieves a desired goal \cite{minsky1988society}? Conventional approaches to this problem, including forward planning and brute-force search, quickly become computationally intractable as the dimensionality of the state-action space increases. This phenomenon, known as the \emph{curse of dimensionality} \cite{bellman1966dynamic}, imposes severe limitations on models that rely on exhaustive simulation \cite{silver2017mastering} or uninformed exploration~\cite{russell2016artificial}.

A critical insight, arising from reinforcement learning, particularly from the temporal difference update rule in Q-learning~\cite{watkins1992q}, is the idea of {\em bootstrapping}. Rather than computing the full trajectory to a reward, Q-learning performs a one-step bootstrapped approximation, using a partial estimate of future value to update current action values. This ``half-step down'' trick bypasses the need for full expansion of the future, suggesting that forward search can be circumvented by local, structured inference. We extend this principle into a more general framework: \emph{inverted inference}. Instead of simulating forward from the present, inverted inference begins with the goal and infers {\em backward} the sequence of states and actions most likely to achieve it~\cite{friston2010free, ha2018world}. Such inverted inference is conceptually and structurally analogous to running dynamic programming (DP) backward, whose optimality is guaranteed by Bellman's principle of optimality \cite{bellman1966dynamic}. Formally, we have

\begin{proposition}[Inverted Inference as Dimensionality-Constrained Bootstrapping]
In forward inference, generating trajectories via $p(\Psi | \Phi)$ requires sampling over the full state-action-context space, leading to exponential complexity in $\dim(\Phi)$. 
Inverted inference reverses this direction by conditioning on goals and inferring compatible contexts via $p(\Phi | \Psi)$. If $\Psi$ is structurally sparse and semantically anchored (e.g., as a goal proxy), then we have
$\dim\left(\operatorname{supp}(p(\Phi | \Psi))\right) \ll \dim(\Phi)$,
transforming the inference problem into a constrained search over a goal-consistent submanifold.
\end{proposition}

\noindent\textbf{Remark:}
In addition to Q-learning, inverted inference is conceptually analogous to the principle of delayed decision in motor control \cite{cisek2005neural}, where the motor system prepares multiple potential actions in parallel but delays commitment to a specific motor plan until more evidence accumulates \cite{todorov2002optimal}.
Similar to postponing action until ambiguity resolves, inverted inference breaks the curse of dimensionality by transforming unconstrained sampling into structure-seeded cycle completion: $\Psi \rightarrow \Phi \rightarrow \Psi$,
where each cycle restricts inference to dimensions aligned with goal semantics.

The above inverted formulation aligns with the optimality principle of dynamic programming, which asserts that optimal substructure enables backward decomposition of global objectives~\cite{bellman1966dynamic}. When paired with a generative world model, goal-directed simulation becomes a powerful means of solving the search problem. It transforms the problem from open-ended expansion to a constrained inference process, where trajectories are sampled not from the full state space but from low-dimensional manifolds anchored on internally specified goals. In doing so, it turns a computational liability into an advantage: high-dimensional latent spaces offer the expressive capacity to encode rich goal hierarchies~\cite{ha2018world} and enable compositional planning~\cite{hafner2019dream}. Therefore, we have the following.

\begin{corollary}[Cycle Formation Enables Dimensionality Reduction]
Under CCUP, the agent alternates between bottom-up contextual disambiguation and top-down content reconstruction by forming inference cycles:
$\Phi \rightarrow \Psi \rightarrow \Phi$.
When initialized by goal-seeded content $\Psi$, inverted inference $p(\Phi | \Psi)$ restricts the search space to context states compatible with $\Psi$. Completing the cycle via forward mapping $p(\Psi | \Phi)$ reinforces consistency, forming a low-dimensional attractor subspace:
$D_{\mathrm{KL}}\big(p(\Phi | \Psi) \, \| \, p_{\text{cycle}}(\Phi)\big) \leq \varepsilon$,
where $p_{\text{cycle}}(\Phi)$ denotes the fixed point of the cycle. 
\end{corollary}

\noindent\textbf{Remark:}
CCUP-based cycle formation enables \textbf{recursive compression} of inference by iteratively projecting onto goal-consistent manifolds, breaking the curse of dimensionality through asymmetric alignment of context and content. While cycle formation reduces dimensionality by aligning context and content within a self-consistent loop, its true computational power emerges when inference is seeded by internal goals, transforming the cycle into a mechanism for simulating goal-consistent trajectories through inverted inference. To fill this gap, we have the following.

\begin{proposition}[Inverted Inference Enables Goal-Directed Simulation]
Let \( \Psi_{\mathrm{goal}} \) be a low-entropy internal goal representation, and let \( \Phi_{0:T} \) denote a high-entropy latent trajectory over time. The process of goal-directed simulation is equivalent to performing inverted inference:
$p(\Phi_{0:T} | \Psi_{\mathrm{goal}}) \propto p(\Psi_{\mathrm{goal}} | \Phi_{0:T}) \cdot p(\Phi_{0:T})$,
which selects trajectories \( \Phi_{0:T} \) that are most consistent with the internal goal.
This inference minimizes the conditional entropy:
$\min_{p(\Phi_{0:T})} \; H(\Phi_{0:T} | \Psi_{\mathrm{goal}})$,
subject to structural constraints imposed by \( \Psi_{\mathrm{goal}} \), restricting simulation to a goal-consistent manifold \( \mathcal{M}_{\mathrm{goal}} \subset \Phi^T \).
\end{proposition}
\noindent\textbf{Remark:}
Inverted inference provides the computational foundation for goal-directed simulation by transforming the search problem into a structure-constrained entropy minimization process aligned with CCUP.
To further formalize this perspective, inverted inference can be understood as a cognitive analog to the primal-dual method in constrained optimization, where internal goals act as dual variables that shape the feasible inference space through structured constraints.

\subsection{Inverted Inference Meets Primal-Dual}

Inverted inference can be understood as a cognitive analog to the primal-dual method in constrained optimization \cite{boyd2004convex}. In classical optimization, the primal variables represent the original search space, while dual variables enforce structural or feasibility constraints via Lagrange multipliers. The solution emerges from a saddle point of the Lagrangian, balancing the objective with the satisfaction of constraints. Similarly, inverted inference begins not with the raw context (primal state) but with a high-level goal (dual constraint) that restricts the latent search space to a goal-consistent manifold \cite{liu2025safe}. From the CCUP perspective, content (low-entropy goals) guides inference over high-entropy context, minimizing joint uncertainty in a manner analogous to dual-based regularization \cite{chambolle2011first}. This correspondence suggests that goal-anchored simulation is a form of constraint-informed inference, where the dual variable (the goal) guides the search over latent states, ensuring structure before specificity. Inverted inference inherits both the recursive optimality of dynamic programming and the constraint-satisfying structure of primal-dual optimization.

\begin{theorem}[Unified View: Primal-Dual and Inverted Inference]
Let \( x \in \mathbb{R}^n \) denote high-dimensional search variables, and \( g(x) = 0 \) encode structural constraints. Let \( \Phi_{\text{goal}} \) denote a low-entropy internal goal that defines a constraint manifold \( \mathcal{M}_{\text{goal}} \).
Then, solving the constrained optimization problem:
$\min_{x} \; f(x)  \text{s.t.}  g(x, \Phi_{\text{goal}}) = 0$
via its Lagrangian:
$\mathcal{L}(x, \lambda) = f(x) + \lambda^\top g(x, \Phi_{\text{goal}})$
is equivalent to performing inverted inference under CCUP, where:
$x \sim p(x | \Phi_{\text{goal}})$
and the dual variable \( \lambda \) enforces consistency between high-entropy context and low-entropy goal (content). The solution lies on a goal-constrained manifold \( \mathcal{M}_{\text{goal}} \subset \mathbb{R}^n \), avoiding exhaustive search and breaking the curse of dimensionality.
\end{theorem}

\noindent\textbf{Remark:} Building on this unified perspective, we now show that inverted inference can be rigorously formulated as a primal-dual entropy minimization problem, in which the alignment between content and context emerges from the interplay between forward (primal) and backward (dual) inference paths.

\begin{theorem}[Inverted Inference as a Primal-Dual Entropy Minimization]
Let $\Phi_{0:T}$ denote a latent trajectory, and let $\Phi_{\text{goal}}$ be a low-entropy terminal state serving as a goal. Let $\mathcal{M}_{\text{goal}} \subseteq \Phi^T$ denote the manifold of trajectories consistent with $\Phi_{\text{goal}}$. The inverted inference problem
$\min_{p(\Phi_{0:T})} \; H(\Phi_{0:T} | \Phi_{\text{goal}})$
subject to the constraint that $p(\Phi_{0:T}) \in \mathcal{M}_{\text{goal}}$
is equivalent to solving the saddle point of the Lagrangian
$\mathcal{L}(p, \lambda) = H(\Phi_{0:T}) + \lambda \cdot \mathbb{E}_{p} \left[ \mathcal{C}(\Phi_{0:T}, \Phi_{\text{goal}}) \right]$,
where $\mathcal{C}$ is a divergence or constraint penalty that enforces goal consistency. The dual variable $\lambda$ enforces structural alignment between the latent trajectory and the goal.
\end{theorem}

\noindent\textbf{Remark:} Inverted inference minimizes conditional uncertainty over latent states (the primal objective), while simultaneously satisfying goal-based structural constraints (the dual coupling), analogous to the primal-dual solution in constrained variational optimization. To deepen the computational interpretation of inverted inference, we now show that its primal-dual entropy minimization form naturally gives rise to a duality with optimal transport, where inference operates as a structured coupling between high-entropy contexts and low-entropy goals.

\begin{corollary}[Inverted Inference and OT Duality]
Let \( \Phi_{0:T} \sim p \) denote a latent trajectory distribution, and let \( \Phi_{\text{goal}} \sim q \) denote a low-entropy goal distribution. Suppose the constraint penalty \( \mathcal{C}(\Phi_{0:T}, \Phi_{\text{goal}}) \) is a transport cost \( c(\Phi_T, \Phi_{\text{goal}}) \) applied to the endpoint \( \Phi_T \). Then the Lagrangian:
$\mathcal{L}(p, \lambda) = H(\Phi_{0:T}) + \lambda \cdot \mathbb{E}_{p} \left[ c(\Phi_T, \Phi_{\text{goal}}) \right]$
reduces to the entropy-regularized OT objective between the trajectory endpoint and the goal distribution:
$\min_{\gamma \in \Pi(p_T, q)} \; \mathbb{E}_{\gamma}[c(\Phi_T, \Phi_{\text{goal}})] - \varepsilon H(\gamma)$, where \( p_T \) is the marginal of \( p(\Phi_{0:T}) \) at \( \Phi_T \), \( q = \delta_{\Phi_{\text{goal}}} \) is a goal-anchored content distribution and \( \Pi(p_T, q) \) denotes couplings with correct marginals.
\end{corollary}

\noindent\textbf{Remark:} Inverted inference under CCUP is equivalent to solving a dual form of entropy-regularized optimal transport, where the \textbf{primal} distribution over trajectories seeks to minimize conditional entropy and the \textbf{dual} enforces alignment of trajectory endpoints with the low-entropy goal via OT. This formulation reveals that CCUP implements an {\em information-constrained transport of inference mass} from high-entropy context to low-entropy content.
While optimal transport duality reveals how inverted inference aligns high-entropy trajectories with low-entropy goals through entropy-regularized couplings, real-world cognition rarely operates at a single level of abstraction. Instead, biological agents must coordinate inference across nested temporal and semantic hierarchies, from immediate sensorimotor contingencies to abstract, long-range goals. To address this, we now extend the framework of inverted inference to a hierarchical setting in space and time, where each layer performs goal-directed simulation by conditioning on higher-level priors while supplying structure to lower levels. This recursive composition enables the system to perform efficient planning and generalization in high-dimensional spaces, transforming inverted inference into a scalable simulation engine. We study recursive temporal bootstrapping next.

\section{Path-dependent Optimization via Recursive Temporal Bootstrapping}

%\subsection{Inverted Inference as Path-Dependent Optimization}

In classical optimization and inference frameworks, agents search or update over a high-dimensional space of possible states, often incurring exponential computational cost as dimensionality increases. This is the hallmark of the \emph{curse of dimensionality}. However, biological and cognitive systems appear to circumvent this issue routinely, achieving efficient decision-making and generalization in complex environments (so-called \textbf{slow thinking} \cite{kahneman2011thinking}). We argue that this efficiency arises from the interplay between \emph{inverted inference} and \emph{path-dependent optimization}, which together transform a high-dimensional search problem into a constrained trajectory over low-dimensional manifolds.
Rather than mapping from high-entropy context to low-entropy content, the agent begins with an internal representation $\Phi$ (e.g., a goal) and infers a consistent generative context $\Psi$ via $P(\Psi | \Phi)$. This directionally inverted approach acts as a dual constraint, restricting the feasible region of the state space \emph{before} search begins.

Crucially, this leads to path-dependent optimization. Each intermediate state along the inference trajectory depends on the preceding steps, forming a temporally structured path through a constrained subspace instead of the full domain. The result is a dynamically pruned manifold:
$\mathcal{M}_\Phi := \left\{ \Psi \in \mathcal{C} \; \middle| \; \exists \text{ path } \{\Psi_t\}_{t=0}^T \text{ s.t. } \Psi_T = \Psi, \; \Psi_t \sim P(\Psi_{t+1} | \Phi) \right\}.$
Inference is thereby restricted not to the entire context space $\mathcal{C}$, but to the structured manifold $\mathcal{M}_\Phi$, which is seeded by the goal and evolved over time through consistent updates. This strategy breaks the curse of dimensionality by replacing global search with goal-directed navigation on a context-conditioned submanifold, which we formalize below.

\begin{theorem}[Inverted Inference Induces Path-Dependent Optimization]
Let $\Phi$ be a low-entropy content variable encoding internal goals, and $\Psi \in \mathcal{C}$ be a high-entropy context variable. Assume inference proceeds via a generative model $P(\Psi | \Phi)$, and let $\mathcal{M}_\Phi$ be the manifold of context states reachable from $\Phi$ via consistent sampling trajectories. Then 1) inference over $\mathcal{M}_\Phi$ is path-dependent: each $\Psi_t$ depends on the trajectory history $\{\Psi_0, \dots, \Psi_{t-1}\}$; 2) the effective dimensionality of $\mathcal{M}_\Phi$ satisfies:
$\dim(\mathcal{M}_\Phi) \ll \dim(\mathcal{C})$, resulting in exponential reduction of the search space; 3) the induced optimization problem $\min_{\Psi \in \mathcal{M}_\Phi} c(\Psi, \Phi)$ is tractable under conditions of local convexity and bounded curvature of $\mathcal{M}_\Phi$.
\end{theorem}

%\subsection{Recursive Temporal Bootstrapping via Chain Memory}
\noindent\textbf{Remark:} The proof of the above theorem is referred to the Appendix.
While inverted inference enables path-dependent optimization by constraining inference to goal-consistent trajectories, its full expressive power emerges when extended across time through recursive temporal bootstrapping, wherein each inference cycle seeds the next via a memory chain that encodes structured dependencies. Next, we first define memory chain as a hierarchical extension of asymmetric cycle in the temporal domain, and then show how memory chain enables recursive structure refinement and hierarchical goal simulation.

\begin{definition}[Memory Chain]
Let $\{Z_t\}_{t=1}^T$ be a sequence of latent variables representing internal memory states indexed over discrete time steps $t = 1, \dots, T$. A \emph{memory chain} is a structured trajectory through latent space such that:

\begin{enumerate}
    \item Each $Z_t$ is conditionally dependent on both the preceding state $Z_{t-1}$ and an associated context variable $X_t$, forming the joint model:
    $p(Z_{1:T}, X_{1:T}) = p(Z_1) \prod_{t=2}^T p(Z_t | Z_{t-1}, X_t)$,
       
    \item The inference over the chain involves iterative bidirectional updates:    
   $q(Z_t) \propto \psi_t(Z_t) \prod_{s \in \mathcal{N}(t)} \phi_{s \to t}(Z_t),$    
    where $\psi_t$ denotes the local potential incorporating contextual evidence, and $\phi_{s \to t}$ are global messages passed from neighboring latent states $Z_s$.
    
    \item The memory chain supports approximate inference by minimizing a variational free energy objective over time:
    $\mathcal{F}[q] = \sum_{t=1}^T \mathbb{E}_{q(Z_t)}[-\log p(X_t | Z_t)] + \text{D}_{KL}(q(Z_{1:T}) \Vert p(Z_{1:T})).$
    
\end{enumerate}

\end{definition}  

\noindent\textbf{Remark:} A memory chain represents a temporally extended, entropy-constrained trajectory of internal states, where each link in the chain performs partial cycle completion between content and context. It generalizes localized cycle-based inference into temporal alignment across multiple levels of abstraction. %Several empirical observations support the biological plausibility of the memory chain. For example, the phase of sloped recession refers to the systematic advancement (precession) of spike timing relative to theta phase as an animal traverses a spatial field (pp. 320, \cite{buzsaki2006rhythms}). The slope of this phase shift encodes spatial or episodic structure and reflects intrinsic oscillatory dynamics and network interactions. The temporal receptive window of neural circuits increases hierarchically along the sensory-to-association cortex axis, enabling cumulative integration of past information over increasing timescales \cite{hasson2015hierarchical}. 
Heuristically, local potentials (context $\psi$) constrain immediate inference, while global potentials (content $\phi$) represent stable representations or structured priors embedded into the memory chain. A crucial missing link is the bridge between local and global potentials, known as cloning \cite{george2021clone}. 

\noindent\textbf{Cloning mechanism and chain formation.} 
The CCUP framework posits that inference operates through cycles that iteratively align high-entropy context variables \( \Psi \) with low-entropy content variables \( \Phi \), minimizing joint uncertainty through bidirectional inference. While individual cycles resolve local ambiguities between context and content, intelligent cognition requires coherence across temporally extended experience.
We propose that such global coherence arises through \emph{chain formation}, defined as the structured \emph{concatenation of inference cycles}:
$(\Psi_1 \leftrightarrow \Phi_1) \rightarrow (\Psi_2 \leftrightarrow \Phi_2) \rightarrow \cdots \rightarrow (\Psi_n \leftrightarrow \Phi_n)$
Each link in the chain corresponds to a local inference cycle, where \( \Phi_k \) is inferred from \( \Psi_k \) and reused or refined in \( \Psi_{k+1} \), possibly via a {\em cloning mechanism} \cite{george2021clone} that allows semantic continuity across varying contexts.

%\paragraph{Cloning as a Bridge Between Cycles.}  
To ensure continuity and generalization, a cloned representation of content \( \Phi_k \) is propagated forward, seeding the next cycle by anchoring it to previously stabilized latent variables. This enables the system to bootstrap learning across steps, supporting goal persistence, memory chaining, and temporal abstraction.
In the temporal domain, cloning supports memory consolidation, where episodic events are linked via shared content variables, forming a coherent narrative structure. In either case, chain formation implements a form of \emph{latent navigation} \cite{ho2022classifierfree,locatello2019challenging}, where the agent traverses an abstract representational space through sequentially coupled inference cycles.
While path-dependent optimization constrains inference to a goal-conditioned manifold, its full potential is realized when this process is recursively extended across time through \emph{temporal bootstrapping}. In this generalization, each completed inference cycle does not merely yield an isolated solution but produces a new constraint, stored in \emph{chain memory} that conditions subsequent cycles. This transforms inference from a single loop to a temporally structured chain of cycles, where each iteration refines both the representational space and the trajectory of future inference.

Let each inverted inference cycle produce a tuple $(\Psi^{(t)}, \Phi^{(t)})$, where $\Psi^{(t)}$ is a disambiguated context and $\Phi^{(t)}$ is its corresponding content. The output $\Phi^{(t)}$ becomes the structural prior (or dual constraint) for the next cycle:
$\Phi^{(t+1)} := \text{Update}(\Phi^{(t)}, \Psi^{(t)}),  \Psi^{(t+1)} \sim P(\Psi | \Phi^{(t+1)})$.
This defines a recursive bootstrapping rule, where inference does not merely converge on a point, but unfolds as a temporally coherent chain:
$(\Psi^{(0)}, \Phi^{(0)}) \rightarrow (\Psi^{(1)}, \Phi^{(1)}) \rightarrow \dots \rightarrow (\Psi^{(T)}, \Phi^{(T)})$.
Each pair in this sequence is stored in a dynamic memory structure we call the \emph{memory chain}:
$\mathcal{M}_{\text{chain}} = \left\{ (\Psi^{(t)}, \Phi^{(t)}) \right\}_{t=0}^T$.
This memory supports counterfactual simulation, long-range planning, and hierarchical goal decomposition. Unlike static representations, memory chain bootstraps structure and specificity across time, allowing the agent to refine its representational manifold with each cycle while reusing prior solutions as new constraints. That is

\begin{proposition}[Memory Chain Enables Recursive Structure Refinement]
Under the conditions of Theorem 6, let each inference cycle update the content variable via a temporally recursive rule:
$\Phi^{(t+1)} := f(\Phi^{(t)}, \Psi^{(t)})$,
where $f$ is a structure refinement function. Then we have

\begin{enumerate}
    \item The inference manifold $\mathcal{M}_{\Phi^{(t)}}$ becomes progressively specialized with each cycle:
    $\mathcal{M}_{\Phi^{(t+1)}} \subseteq \mathcal{M}_{\Phi^{(t)}}$.

    \item The effective optimization path becomes increasingly constrained:
    $\bigcap_{t=0}^T \mathcal{M}_{\Phi^{(t)}} \subseteq \mathcal{C}$,
    enabling hierarchical pruning of the search space.

    \item The chain memory $\mathcal{M}_{\text{chain}}$ encodes a temporally ordered sequence of structure–specificity refinements, supporting generalization across tasks and time.
\end{enumerate}
\end{proposition}

\noindent\textbf{Remark:} A k=sketch of the proof can be found in Appendix. Phase resetting \cite{buzsaki2006rhythms} provides the neurophysiological substrate for initiating new inference cycles within the memory chain by acting as the temporal boundary condition that gates when and how recursive structure refinement occurs. Together, memory chain and phase resetting implement a temporally grounded form of structure learning, where cognitive architecture is rhythmically re-aligned with goals and observations. An important new insight offered by this realignment is the construction of temporally unfolding goal hierarchies through
structure-preserving bootstrapping \cite{zhao2024model}. We conclude this section with the following proposition (proof skipped).

\begin{proposition}[Recursive Temporal Bootstrapping Enables Hierarchical Goal Simulation]
Let \( \Psi_{\mathrm{goal}} \) be a high-level internal goal and let \( \{(\Phi_t^{(n)}, \Psi_t^{(n)})\}_{n=0}^{N} \) denote a memory chain of inference cycles over time. Recursive temporal bootstrapping defines a sequence of subgoal inferences:
$\Psi^{(n)}_t \sim p(\Psi_t^{(n)} | \Phi_t^{(n-1)}, \Psi_t^{(n-1)}),  \text{for } n = 1, \dots, N$,
where each subgoal \( \Psi^{(n)}_t \) serves as a goal anchor for the next inference cycle.
This recursive structure enables hierarchical goal decomposition over time:
$\Psi_{\mathrm{goal}}=\Psi^{(0)}_t \rightarrow \Psi^{(1)}_t \rightarrow \Psi^{(2)}_t \rightarrow \cdots \rightarrow \Psi^{(N)}_t$,
such that each subgoal refines and temporally localizes the overarching plan.
\end{proposition}

\noindent\textbf{Remark:} Memory chains under recursive inverted inference support both simulation of trajectories and construction of temporally unfolding goal hierarchies through structure-preserving bootstrapping \cite{correa2025exploring}. This recursive formulation offers a computational model for hierarchical planning without the need for predefined goal trees. Subgoals are not statically declared but emerge as information-theoretically optimal anchors in the temporal simulation chain \cite{kaplan2018planning}. Such a mechanism supports flexible re-planning, nested intentions, and dynamic compositionality—core features of intelligent behavior observed in biological agents \cite{goyal2019infobot}. Moreover, this model provides a natural explanation for the temporal organization of goal-directed cognition, including the role of episodic memory, prospective simulation, and cognitive control. In this view, subgoals are stabilized as phase-locked attractors within the memory chain, enabling dynamic but coherent goal unfolding across time.

\section{Layered Primal-Dual Inference for Hierarchical Spatial Bootstrapping}

Real-world cognition requires agents to reason across multiple levels of abstraction in space, from low-level sensorimotor interactions to high-level, spatially clustered objects. It has been hypothesized that the brain's hierarchical organization reflects the nested structure of the physical world itself \cite{hawkins2021thousand}, enabling efficient abstraction and generalization. In particular, the neocortex exhibits a small-world architecture optimized to maximize information flow by recursively breaking the information bottleneck (IB) across spatial scales \cite{buzsaki2006rhythms}. The principle of dynamic alignment through context disambiguation and reconstruction naturally extends across hierarchical levels of cognitive inference \cite{friston2008hierarchical}. In hierarchical systems, whether biological or artificial, lower layers perform rapid (so-called \textbf{fast thinking} \cite{kahneman2011thinking}), fine-grained disambiguation of sensory input, while higher layers accumulate more abstract, temporally extended contextual information. Unlike temporal bootstrapping, which encodes causal dependency, spatial bootstrapping encodes structural dependency \cite{kingma2014auto}.

Crucially, these cycles are not isolated but are recursively coupled: the disambiguated content from one level serves as contextual input for the next, and higher-level abstractions guide lower-level predictions \cite{dayan1995helmholtz}. This nested structure allows uncertainty to be progressively redistributed across layers, enabling both specificity at the sensory periphery and generalization at abstract levels. The result is a scalable architecture in which inference converges not only within each level but also across the hierarchy \cite{friston2008hierarchical}, forming a cascade of dynamically aligned cycles in both space and time. Such multiscale coordination explains how complex cognitive systems maximize the information flow despite the recurrence of IB across scales, which is consistent with the idea underlying constructal theory \cite{bejan2008design}. Formally, we extend the definition of dynamic alignment as follows.

\begin{definition}[Hierarchical Extension of Dynamic Alignment]
Let a hierarchical system be composed of \( L \) levels of abstraction, each with latent representations \( Z^{(1)}, \dots, Z^{(L)}=\Phi \), and contextual observations \( \Psi = Z^{(0)} \). Under CCUP, each level \( \ell \in \{1, \dots, L\} \) performs a local inference cycle:
$Z^{(\ell - 1)} \xrightarrow{\text{encoding}} Z^{(\ell)} \xrightarrow{\text{decoding}} \hat{Z}^{(\ell - 1)} \xrightarrow{\text{update}} Z^{(\ell)\prime}$
such that: 

1) {Bottom-up encoding:} \( Z^{(\ell)} \) is initialized by an anchor seed as a low-entropy distribution $p(z_0^{(l)}|Z^{(l-1)})$, enforcing strong content anchoring; 

2) {Top-down decoding:} A manifold constrained projection process \( \mathcal{D}^{(\ell)} \) unfolds \( z^{(\ell)}_0 \) into context-conditioned predictions: $\hat{Z}^{(\ell - 1)} = \mathcal{D}^{(\ell)}(z^{(\ell)}_0; Z^{(\ell + 1)})$ guided by higher-level constraints \( Z^{(\ell + 1)} \); 

3) {Cycle-consistent feedback:} Discrepancy between \( Z^{(\ell - 1)} \) and \( \hat{Z}^{(\ell - 1)} \) is measured: 
$\delta^{(\ell)} = d\left(Z^{(\ell - 1)}, \hat{Z}^{(\ell - 1)}\right),$ 
and used to refine \( Z^{(\ell)} \) through a self-supervised update: $Z^{(\ell)\prime} = Z^{(\ell)} - \eta \nabla_{Z^{(\ell)}} \delta^{(\ell)}.$

If each local cycle minimizes \( \delta^{(\ell)} \leq \epsilon \), then the full system converges toward a globally cycle-consistent hierarchy, minimizing conditional entropy at each level and maximizing mutual information:
$I(Z^{(\ell - 1)}; Z^{(\ell)}) \approx H(Z^{(\ell - 1)}).$
This resolves the IB across levels by dynamically aligning content and context representations under delta-seeded inference.
\end{definition}

%\section{Hierarchical Spatial Bootstrapping for Fast Object Recognition}

Inverted inference provides a powerful framework for simulating goal-consistent latent causes from abstract, low-entropy goals. When extended spatially, this process enables the rapid identification of complex objects by leveraging the structural hierarchy of feature representations in the visual system. Here, we introduce \textit{hierarchical spatial bootstrapping} as a mechanism for fast object recognition and show how it naturally admits an interpretation of layered primal-dual inference.

\noindent\textbf{Inference as Coarse-to-Fine Hypothesis Refinement}.
Object recognition is classically treated as a feedforward process, wherein low-level features (e.g., edges, textures) are aggregated into mid-level parts and eventually recognized as whole objects. However, this purely bottom-up view fails to explain the remarkable speed and robustness of biological vision, particularly under occlusion, ambiguity, or incomplete information.
Hierarchical spatial bootstrapping reframes recognition as an inverted inference process:
\begin{itemize}
    \item A high-level goal or category prior (e.g., ``cup'') initiates a top-down search for compatible low-level evidence.
    \item Each level of the spatial hierarchy performs entropy minimization subject to structural alignment constraints with its parent.
    \item Recognition is achieved when top-down priors and bottom-up features are dynamically aligned through cycle completion.
\end{itemize}

%\textbf{Formal Proposition: Goal-Conditioned Spatial Inference}
This coarse-to-fine refinement process forms the basis of hierarchical spatial bootstrapping, which enables fast object recognition by rapidly aligning top-down semantic priors with bottom-up sensory evidence across multiple levels of abstraction. Formally, we have

\begin{proposition}[Hierarchical Spatial Bootstrapping Enables Fast Object Recognition]
Let \( \Psi_{\mathrm{object}} \) be a high-level goal or category prior, and let \( \{\Phi^{(\ell)}\}_{\ell=0}^{L} \) be a spatial inference hierarchy. Then object recognition proceeds via inverted inference:
$p(\Phi^{(0)} | \Psi_{\mathrm{object}}) \sim \arg\min_{p} H(\Phi^{(0)} | \Psi_{\mathrm{object}})$
subject to recursive constraints:
$\mathbb{E}_{p^{(\ell)}} \left[ \mathcal{C}^{(\ell)}(\Phi^{(\ell)}, \Phi^{(\ell+1)}) \right] \leq \varepsilon,  \forall \ell$.
\end{proposition}

\noindent\textbf{Remark:} This hierarchical bootstrapping restricts inference to a goal-consistent manifold in spatial feature space, enabling rapid cycle completion and pattern recognition.
Therefore, inverted inference with hierarchical spatial bootstrapping supports ``fast thinking'' by efficiently aligning sensory input with top-down semantic structure.
\begin{comment}
\noindent\textbf{Biological Interpretation:}
Neuroscientific evidence supports the idea that the brain uses hierarchical feedback to accelerate and disambiguate object recognition. Specifically:
\begin{itemize}
    \item The \textbf{ventral visual stream} exhibits a hierarchical structure from V1 through inferotemporal cortex (IT), with increasingly abstract and invariant representations.
    \item \textbf{Top-down signals} from higher visual areas (e.g., IT, prefrontal cortex) arrive within 100-200 ms and modulate lower-level feature activity, suggesting fast semantic bootstrapping.
    \item \textbf{Predictive coding} models \cite{rao1999predictive,friston2005theory} show that recognition occurs through error minimization between expected and observed features, a direct implementation of CCUP cycles.
    \item \textbf{Fast object recognition} even under partial occlusion \cite{grill2000dynamics} supports the view that structure is filled in rapidly via top-down inference, rather than constructed purely bottom-up.
\end{itemize}
\end{comment}
In this light, hierarchical spatial bootstrapping offers a unified computational and biological explanation for ``fast thinking'' in object recognition \cite{kahneman2011thinking}. Rather than passively constructing representations, the brain actively simulates, tests, and updates hypotheses in a coarse-to-fine manner, guided by semantically grounded goals and contextual priors.

\noindent\textbf{Layered primal-dual interpretation.}
This framework implements a \textit{layered primal-dual architecture} in which higher layers provide low-entropy structural priors, and lower layers infer fine-grained content consistent with those constraints. Each layer seeks to minimize uncertainty (entropy) in its local latent representation while maintaining semantic alignment with higher-level goals and supporting inference for lower levels. These nested cycles form a \emph{multiscale generalization} of dynamic alignment, enabling the system to simulate goal-directed trajectories across levels while preserving cross-scale consistency. In parallel to Proposition 4 for temporal bootstrapping, we have the following result for spatial cognition.

\begin{proposition}[Hierarchical Goal-Directed Simulation as Layered Primal-Dual Inference]
Let $\{\Phi^{(\ell)}\}_{\ell=0}^L$ denote latent variables across $L$ hierarchical layers, where $\Phi^{(L)}$ represents a top-level goal. Assume each layer $\ell$ solves a local primal-dual problem:
$\min_{p^{(\ell)}(\Phi^{(\ell)})} \; H(\Phi^{(\ell)} | \Phi^{(\ell+1)}) ~\text{s. t.}~ \mathbb{E}_{p^{(\ell)}} \left[ \mathcal{C}^{(\ell)}(\Phi^{(\ell)}, \Phi^{(\ell+1)}) \right] \leq \varepsilon$.
Then the entire hierarchy implements a structured inference process that minimizes the global objective:
\begin{align}
    \min_{\{p^{(\ell)}\}} \; \sum_{\ell=0}^{L-1} \left[ H(\Phi^{(\ell)} | \Phi^{(\ell+1)}) + \lambda^{(\ell)} \cdot \mathbb{E}_{p^{(\ell)}} \left[ \mathcal{C}^{(\ell)}(\Phi^{(\ell)}, 
\Phi^{(\ell+1)}) \right] \right],
\end{align}
where each dual variable $\lambda^{(\ell)}$ enforces semantic alignment between adjacent layers.
\end{proposition}
\noindent\textbf{Remark:} 
This layered primal-dual structure enables hierarchical goal-directed simulation, where high-level abstract goals propagate downward as structural priors
and lower layers simulate compatible trajectories by minimizing conditional entropy.
Cycles across layers align context and content dynamically, forming a multiscale extension of inverted inference in the spatial domain. To shed light on how such hierarchical spatial bootstrapping supports fast thinking, we borrow insight from navigating in a small world \cite{kleinberg2000navigation} and show how to break the IB recurrent across different scales.

%\subsection{Cloning Expansion Meets Manifold Projection}

\noindent\textbf{Cloning expansion for scalable OT.}
A complementary intuition for resolving the IB in a scalable OT solution arises from the theory of hierarchical navigable small-world (HNSW) networks \cite{malkov2018efficient}. These networks enable efficient search and routing by embedding latent representations across nested spatial or semantic scales. Under the CCUP framework, the asymmetry of directional information loss implies that the IB reappears recursively: disambiguating local context while preserving global structure requires resolving uncertainty not just once, but at every scale of inference \cite{friston2008hierarchical}. To address this, a \emph{hierarchical extension of the cloning trick} is required. At each level of abstraction, cloned latent representations can capture context-specific variants of globally relevant content, allowing inference cycles to operate independently yet coherently across the hierarchy \cite{bengio2013representation, esmaeili2019structured}. This hierarchical cloning ensures that latent structure remains navigable and distinct, preventing aliasing between similar substructures encountered at different scales. As a result, the agent's internal representation mirrors the nested, compositional structure of the external world. In this view, hierarchical cloning enables scalable generalization without sacrificing specificity and supports efficient inference and goal-directed behavior across the full spectrum of context, from local sensory details to abstract task space.

\noindent\textbf{Manifold-constrained Projection.} Having established how hierarchical cloning preserves the separability and navigability of latent structure across scales, we now turn to the complementary process of reconstruction—specifically, how top-down context reconstruction generatively expands compact content representations into plausible contextual configurations.
%Top-down context reconstruction refers to the generative process of reconstructing plausible contextual configurations from compact content representations. 
Rather than exploring the full high-dimensional space of possible contexts, this reconstruction under inverted inference is constrained to unfold along a learned low-dimensional manifold shaped by prior experience and structural regularities. These manifolds encode the lawful variations of context conditioned on latent content, such as typical spatial arrangements and semantic associations, thereby guiding reconstruction toward coherent and generalizable interpretations \cite{higgins2017beta}. 

Within predictive coding architectures, manifold constraint is implemented by top-down predictions that are not arbitrary but restricted to lie on trajectories supported by generative priors \cite{chung2022improving}. This manifold constraint serves two critical functions: 1) it regularizes inference by preventing overfitting to unstructured contextual variations; and 2) it promotes abstraction by aligning new observations with previously learned contextual patterns. In this way, manifold-constrained reconstruction ensures that the inverse mapping from content to context remains both efficient and meaningful, enabling robust generalization and the contextual reuse of compact representations. This view of inference as a manifold-constrained projection naturally leads to efficient implementation through sublinear goal retrieval \cite{malkov2018efficient}, where HNSW-accelerated inverted inference enables rapid access to goal-consistent states by navigating the structured latent manifold. That is,

\begin{corollary}[Sublinear Goal Retrieval via HNSW-Accelerated Inverted Inference]
Let \( \mathcal{X} \subset \mathbb{R}^d \) be a high-dimensional latent space organized by a hierarchical inverted inference model. Suppose an HNSW graph is built over \( \mathcal{X} \) with semantic locality preserved.
Then, conditioned on a low-entropy goal \( \Psi_{\text{goal}} \), the retrieval of compatible latent states \( \Phi \in \mathcal{M}_{\text{goal}} \) can be performed in sublinear time:
$\Phi^\ast = \arg\max_{\Phi \in \mathcal{X}} p(\Phi | \Psi_{\text{goal}}) \text{via HNSW} \Rightarrow  O(\log n)$.
Therefore, HNSW-accelerated search enables biologically and computationally plausible scaling of spatially hierarchical inverted inference.
\end{corollary}

\noindent\textbf{Remark:} The corollary above demonstrates that HNSW-based retrieval within a goal-conditioned manifold \( \mathcal{M}_{\text{goal}} \subset \mathcal{X} \) enables sublinear-time inference by leveraging semantic locality and hierarchical graph topology. This mirrors a foundational biological principle: the neocortex exhibits a fractal-like, small-world organization that supports rapid, context-sensitive inference over high-dimensional representations. According to \emph{constructal theory} \cite{bejan2000shape}, natural systems evolve flow architectures that maximize access to essential resources, such as energy, information, or goals, through scale-invariant, branching structures. Within this framework, the hierarchical inverted inference model, augmented by HNSW search, formalizes a mechanism of efficient information flow: low-entropy goals \( \Psi_{\text{goal}} \) act as structural anchors that constrain the search space, while the navigable small-world geometry ensures that relevant content \( \Phi \) is retrieved with minimal computational cost. This dual optimization of structure and function suggests that \emph{inverted inference, when embedded in fractal-like and constructal architectures, enables scalable cognition via goal-directed, structure-preserving search}.

\section{Conclusions}

The brain does not optimize in an empty space. It optimizes over representations, which are themselves constraints that make learning, planning, and inference tractable.
This work presents a unified framework for cognition grounded in CCUP, where inference is framed as the dynamic minimization of conditional entropy between high-entropy contexts and low-entropy content representations. Our main contributions advance the theory and implementation of structured inference in four key directions.

First, we formalized cognition as a process of \textit{dynamic alignment} between context and content. By iterating cycles of bottom-up disambiguation and top-down reconstruction, CCUP transforms inference into convergence on a constraint manifold, eliminating the information bottleneck in optimal transport. This principled cycle formation allows the agent to iteratively reduce uncertainty and resolve ambiguity.

Second, we showed that \textit{inverted inference} enables goal-directed simulation by reversing the causal arrow of inference. Inspired by the structure-before-specificity principle, we constructed an asymmetric cycle in which internal goals act as dual constraints. This formulation closes the duality gap in constrained optimization: when content and context are fully aligned, the system achieves a state of zero divergence, indicating complete semantic convergence.

Third, we extended this framework temporally via \textit{recursive bootstrapping}, wherein each inference cycle dynamically updates the constraint manifold for the next. This gives rise to a form of \textit{path-dependent optimization}, in which memory chains encode temporally structured inference trajectories. This recursive organization bridges cognition with planning and imagination, supporting a dynamic theory of slow thinking grounded in cycle completion rather than static convergence.

Finally, we generalized the CCUP framework spatially through \textit{hierarchical spatial bootstrapping}, connecting it to small-world topologies via the HNSW model. This spatial extension enables fast and scalable goal retrieval through layered primal-dual inference. By embedding the goal-consistent manifold in a navigable small-world graph, HNSW-accelerated inverted inference achieves sublinear retrieval, offering a biologically and computationally plausible model for fast thinking and object recognition.

Together, these contributions provide a comprehensive computational framework in which cognition emerges from the recursive, hierarchical alignment of structure and specificity across time and space. The CCUP-based perspective integrates slow, path-dependent reasoning and fast, goal-anchored recognition into a single inferential architecture grounded in dynamic entropy minimization.

\bibliographystyle{IEEEtran}
\bibliography{reference}

%%%%%%%%%%%%%%%%%%%%%%%%%%%%%%%%%%%%%%%%%%%%%%%%%%%%%%%%%%%%

\appendix
%\section{Appendix}

\noindent\textbf{Proof of Lemma 1}
\label{appendix:A}

\begin{proof}[Proof]
We start from the definition of mutual information:
$I(\Psi; \Phi) = H(\Phi) - H(\Phi | \Psi) = H(\Psi) - H(\Psi | \Phi)$.

Rearranging both expressions:
$H(\Phi | \Psi) = H(\Phi) - I(\Psi; \Phi),  H(\Psi | \Phi) = H(\Psi) - I(\Psi; \Phi)$.

Summing these two expressions:
$H(\Phi | \Psi) + H(\Psi | \Phi) = H(\Phi) + H(\Psi) - 2 I(\Psi; \Phi)$.

Now, using the upper bound on mutual information:
$I(\Psi; \Phi) \leq \min\{H(\Phi), H(\Psi)\}$,

we obtain the lower bound:
$H(\Phi | \Psi) + H(\Psi | \Phi) \geq H(\Phi) + H(\Psi) - 2 \min\{H(\Phi), H(\Psi)\}$.

This simplifies to:
$H(\Phi | \Psi) + H(\Psi | \Phi) \geq |H(\Phi) - H(\Psi)|$.

This bound captures a fundamental implication of CCUP: due to the inherent asymmetry in entropy between context and content, directional uncertainty cannot be jointly minimized beyond the entropy gap. The inequality is tight when mutual information saturates at \( \min\{H(\Phi), H(\Psi)\} \), corresponding to maximal representational alignment.
\end{proof}

\noindent\textbf{Proof of Lemma 2}
\label{appendix:A2}

\begin{proof}
We begin with the joint variational objective under CCUP:
$\min_{p(\Psi, \Phi)} H(\Psi | \Phi) + H(\Phi)$,
which encourages both conditional disambiguation of high-entropy context \( \Psi \) given structured content \( \Phi \) and compression of the content representation \( \Phi \), reflecting its role as a low-entropy semantic anchor.
By the chain rule of entropy, this objective is equivalent to:
$H(\Psi, \Phi) = H(\Psi | \Phi) + H(\Phi)$,
so the problem becomes minimizing the total entropy of the joint distribution over context and content representation, subject to their alignment.

Now consider a generative process in which the contextual variable \( \Psi \) arises from an underlying latent variable \( x \in \mathbb{R}^n \), i.e., \( \Psi = \Psi(x) \) and the representation \( \Phi \) is a function \( g(x) \) that extracts or enforces structure from \( x \).
We now reinterpret the entropy minimization problem as a constrained optimization over latent space \( x \), with \( \Phi \) acting as a constraint on valid latent states. Let \( f(x) \) be a cost or negative log-likelihood function such that \( \Psi \sim p(\Psi(x)) \propto \exp(-f(x)) \). Then the entropy minimization becomes equivalent to:
$\min_{x} f(x)  \text{subject to}  g(x) = \Phi$,
where the constraint enforces that the latent variable \( x \) resides on the manifold \( \mathcal{M}_\Phi = \{x \in \mathbb{R}^n : g(x) = \Phi\} \).
In this setting:
\begin{itemize}
  \item \( x \) is the primal variable over which inference is performed,
  \item \( g(x) = \Phi \) defines a constraint surface shaped by the representation,
  \item \( \Phi \) acts as a dual variable: it restricts the support of \( p(x) \) and regulates the entropy of \( \Psi(x) \).
\end{itemize}
Thus, minimizing \( H(\Psi | \Phi) + H(\Phi) \) is equivalent to optimizing over latent structure with semantic constraints imposed by \( \Phi \). This establishes the equivalence between CCUP inference and constrained optimization, where representations \( \Phi \) act as dual constraints on latent space inference.
\end{proof}

\noindent\textbf{Proof of Theorem 1}
\label{appendix:B}

\begin{proof}
Let $X \sim \mu$, and define $Y := T(X) \sim \nu$. Since $T_{\#}\mu = \nu$, this defines a valid transport plan. Because $T$ is invertible almost everywhere under $\mu$ with inverse $T^{-1}$, the joint distribution $(X, Y)$ induces a deterministic mapping:
$X = T^{-1}(Y),  \mu\text{-almost surely}$.
This implies that $X$ is a deterministic function of $Y$, so the conditional entropy satisfies:
$H(X | Y) = 0$.
By the definition of mutual information:
$I(X; Y) = H(X) - H(X | Y) = H(X)$.
Hence,
$\text{IB Loss} = H(X) - I(X; Y) = H(X) - H(X) = 0$.
Thus, the information bottleneck is resolved.
\end{proof}

\noindent\textbf{Proof of Theorem 2}
\label{appendix:C}

\begin{proof}
We aim to show that under the Context-Content Uncertainty Principle (CCUP), minimizing \( H(\Psi | \Phi) \) to construct a generalizable structure \( \Phi \) prior to minimizing \( H(\Phi | \Psi) \) for content inference leads to lower inference error and more stable representations.

\textbf{Step 1: Entropic Asymmetry under CCUP.} 
By CCUP, context variables \( \Psi \) are high-entropy and ambiguous, while content variables \( \Phi \) are lower-entropy and selectively encoded. That is,
$H(\Psi) \gg H(\Phi)$,
which implies:
$H(\Phi | \Psi) > H(\Psi | \Phi)$.
From the definition of mutual information:
$I(\Psi; \Phi) = H(\Psi) - H(\Psi | \Phi) = H(\Phi) - H(\Phi | \Psi)$,
it follows that minimizing \( H(\Psi | \Phi) \) yields greater gains in mutual information when \( \Phi \) is still underdetermined.

\textbf{Step 2: Generalizable Structure Minimizes Cross-Context Variance.} 
Let \( \Phi^\ast \) be the representation that minimizes expected context entropy across observations:
$\Phi^\ast = \arg\min_{\Phi} \mathbb{E}_{i=1}^N \left[ H(\Psi_i | \Phi) \right]$.
This optimization extracts invariant latent structure across variable contexts. For any new context \( \Psi_j \), this structural prior ensures that:
$\mathbb{E}_j \left[ H(\Psi_j | \Phi^\ast) \right] < \mathbb{E}_j \left[ H(\Psi_j | \Phi_{\text{early}}) \right]$,
where \( \Phi_{\text{early}} \) is a context-specific representation inferred without prior structure, which is more prone to overfitting and aliasing.

\textbf{Step 3: Specificity after Structure Minimizes Inference Error.} 
Let the expected inference loss be:
$\mathcal{L} = \mathbb{E}_j \left[ H(\Phi | \Psi_j) \right]$.
By the chain rule of entropy:
$H(\Phi | \Psi_j) = H(\Phi | \Psi_j, \Phi^\ast) + I(\Phi ; \Phi^\ast | \Psi_j)$.
Since \( \Phi^\ast \) is a learned structural prior that compresses variability across contexts, it reduces the conditional entropy of \( \Phi \) given \( \Psi_j \):
$H(\Phi | \Psi_j, \Phi^\ast) < H(\Phi | \Psi_j)$.
Therefore, conditioning inference on prior structure strictly improves specificity:
$\mathcal{L}_{\text{with structure}} < \mathcal{L}_{\text{without structure}}$.

\textbf{Conclusion.} 
Under CCUP, first minimizing \( H(\Psi | \Phi) \) enables generalization through structure formation, and only then minimizing \( H(\Phi | \Psi) \) enables robust, low-error specificity. Hence, structure must precede specificity for effective inference and memory.
\end{proof}

\noindent\textbf{Proof of Theorem 3}
\label{appendix:D}

\begin{proof}
$(a) \Rightarrow (b)$:  
Cycle consistency implies that forward transport followed by backward reconstruction induces negligible divergence from the original context. This means the solution $\Phi$ respects the constraints imposed by $\lambda$, and thus satisfies the saddle-point condition of the Lagrangian—closing the duality gap.

$(b) \Rightarrow (c)$:  
If the duality gap is zero, the optimal content $\Phi^*$ lies on the manifold $\mathcal{M}_\lambda$, and the coupling $\gamma_\lambda^*$ minimizes the transport cost subject to the constraint $\Phi \in \mathcal{M}_\lambda$. The entropy regularization term \( -\varepsilon H(\gamma) \) enforces smoothness and stability of the map, and the manifold constraint ensures low-dimensionality.

$(c) \Rightarrow (a)$:  
An optimal coupling that respects the constraint manifold and entropy minimization ensures that the transported $\Phi$ can reconstruct $\Psi$ with minimal uncertainty. Hence, cycle-consistent reconstruction is achieved, closing the context–content loop.
Therefore, the asymmetric inference cycle induced by the structure-before-specificity principle results in optimal, low-dimensional alignment between context and content.
\end{proof}

\noindent\textbf{Proof of Theorem 4}
\label{appendix:E}

\begin{proof}
We begin by considering the constrained optimization problem:
$\min_{x} \; f(x)  \text{subject to}  g(x, \Phi_{\text{goal}}) = 0$,
where \( f(x) \) is a cost or loss function defined over high-dimensional variables \( x \in \mathbb{R}^n \), and \( \Phi_{\text{goal}} \) represents a low-entropy latent goal that defines the constraint surface \( \mathcal{M}_{\text{goal}} = \{x \in \mathbb{R}^n : g(x, \Phi_{\text{goal}}) = 0\} \).

Construct the Lagrangian:
$\mathcal{L}(x, \lambda) = f(x) + \lambda^\top g(x, \Phi_{\text{goal}})$,
with dual variables \( \lambda \in \mathbb{R}^m \) enforcing soft constraint satisfaction. The necessary conditions for optimality are given by the Karush-Kuhn-Tucker (KKT) conditions:
\begin{align*}
\nabla_x \mathcal{L}(x, \lambda) &= \nabla f(x) + \nabla_x g(x, \Phi_{\text{goal}})^\top \lambda = 0, \\
g(x, \Phi_{\text{goal}}) &= 0.
\end{align*}

Now interpret the constrained distribution over \( x \) as a posterior:
$p(x | \Phi_{\text{goal}}) \propto \exp(-f(x)) \cdot \delta(g(x, \Phi_{\text{goal}}))$,
where the delta function enforces hard constraint satisfaction. This defines inverted inference conditioned on a low-entropy internal goal: we infer plausible high-dimensional causes \( x \) consistent with the fixed content \( \Phi_{\text{goal}} \).

In this formulation:
- The \textbf{primal variable} \( x \) corresponds to the high-entropy context being inferred.
- The \textbf{dual variable} \( \lambda \) corresponds to an internal constraint enforcer that ensures semantic consistency between content and context.
- The \textbf{goal} \( \Phi_{\text{goal}} \) defines a constraint manifold \( \mathcal{M}_{\text{goal}} \subset \mathbb{R}^n \), drastically reducing the search space.

Hence, solving the optimization problem is equivalent to \textbf{sampling from the posterior} \( p(x I \Phi_{\text{goal}}) \) constrained to the goal-consistent manifold, as dictated by the CCUP framework.
Finally, the dimensionality reduction arises because inference is restricted to the submanifold \( \mathcal{M}_{\text{goal}} \), rather than exploring the full space \( \mathbb{R}^n \), thereby breaking the curse of dimensionality through structure-before-specificity.

\end{proof}

\noindent\textbf{Proof of Theorem 5}
\label{appendix:F}

\begin{proof}
We begin with the formulation of inverted inference as a constrained entropy minimization problem:
$\min_{p(\Phi_{0:T})} \; H(\Phi_{0:T} | \Phi_{\text{goal}})
 \text{subject to}  p(\Phi_{0:T}) \in \mathcal{M}_{\text{goal}}$.
Here, $\Phi_{0:T}$ denotes a trajectory of latent states, and $\Phi_{\text{goal}}$ defines a low-entropy semantic anchor. The constraint set $\mathcal{M}_{\text{goal}}$ comprises all trajectories that terminate in (or are semantically aligned with) the goal.
By the chain rule of entropy:
$H(\Phi_{0:T} I \Phi_{\text{goal}}) = H(\Phi_{0:T}) - I(\Phi_{\text{goal}} ; \Phi_{0:T})$,
minimizing conditional entropy encourages trajectories that are maximally informative about the goal under a structural alignment constraint.

To enforce the constraint $p(\Phi_{0:T}) \in \mathcal{M}_{\text{goal}}$, we relax it via a penalty functional $\mathcal{C}(\Phi_{0:T}, \Phi_{\text{goal}})$ that measures the divergence (e.g., KL, squared error, OT cost) from the goal manifold. The constrained problem becomes a primal-dual variational objective:
$\mathcal{L}(p, \lambda) = H(\Phi_{0:T}) + \lambda \cdot \mathbb{E}_{p} \left[ \mathcal{C}(\Phi_{0:T}, \Phi_{\text{goal}}) \right]$.
The dual variable \( \lambda \) acts as a Lagrange multiplier enforcing **goal-consistent structure** over the latent trajectory. The objective is convex in \( p \) (under standard entropy regularity conditions) and linear in \( \lambda \), yielding a saddle point optimization:
$\min_{p} \max_{\lambda \geq 0} \; \mathcal{L}(p, \lambda)$.

This formulation has a natural interpretation under CCUP:
- The primal term \( H(\Phi_{0:T}) \) encourages exploration over the trajectory space.
- The dual penalty \( \mathbb{E}_p[\mathcal{C}] \) enforces semantic alignment with the low-entropy content anchor \( \Phi_{\text{goal}} \).
- The solution \( p^\ast(\Phi_{0:T}) \) is the distribution over trajectories that is maximally consistent with the goal while being minimally entropic.
Therefore, solving this saddle point problem is equivalent to performing inverted inference as goal-constrained entropy minimization. This avoids brute-force sampling over all trajectories and instead identifies a goal-consistent manifold \( \mathcal{M}_{\text{goal}} \), thus breaking the curse of dimensionality.

\end{proof}

\noindent\textbf{Proof of Theorem 6}
\label{appendix:G}

\begin{proof}
We begin by considering a hierarchy of latent variables $\Phi^{(0)}, \Phi^{(1)}, \dots, \Phi^{(L)}$, where each $\Phi^{(\ell)}$ represents content at abstraction level $\ell$, and $\Phi^{(L)}$ denotes a fixed top-level goal (low-entropy prior). 

For each layer $\ell$, we define a local inference problem:
$\min_{p^{(\ell)}(\Phi^{(\ell)})} \; H(\Phi^{(\ell)} | \Phi^{(\ell+1)})  \text{s.t.}  \mathbb{E}_{p^{(\ell)}} \left[ \mathcal{C}^{(\ell)}(\Phi^{(\ell)}, \Phi^{(\ell+1)}) \right] \leq \varepsilon$,

where $\mathcal{C}^{(\ell)}$ is a cost or divergence that measures semantic misalignment between layer $\ell$ and its parent layer $\ell+1$. Introducing the dual variable $\lambda^{(\ell)}$, we convert the constrained problem into the penalized Lagrangian form:
$\mathcal{L}^{(\ell)} = H(\Phi^{(\ell)} | \Phi^{(\ell+1)}) + \lambda^{(\ell)} \cdot \mathbb{E}_{p^{(\ell)}} \left[ \mathcal{C}^{(\ell)}(\Phi^{(\ell)}, \Phi^{(\ell+1)}) \right]$.

Now, define the **global hierarchical objective** as the sum of all local layer-wise objectives:
$\mathcal{L}_{\text{global}} = \sum_{\ell=0}^{L-1} \mathcal{L}^{(\ell)} = \sum_{\ell=0}^{L-1} \left[ H(\Phi^{(\ell)} | \Phi^{(\ell+1)}) + \lambda^{(\ell)} \cdot \mathbb{E}_{p^{(\ell)}} \left[ \mathcal{C}^{(\ell)}(\Phi^{(\ell)}, \Phi^{(\ell+1)}) \right] \right]$.

Minimizing this global objective with respect to all marginal posteriors $p^{(\ell)}$ defines a structured inference process in which:
- Each layer $\ell$ receives structural priors from $\Phi^{(\ell+1)}$ (top-down).
- It generates content representations $\Phi^{(\ell)}$ (bottom-up) that minimize conditional entropy.
- The constraint penalty $\mathcal{C}^{(\ell)}$ ensures semantic compatibility between adjacent layers.
- Dual variables $\lambda^{(\ell)}$ adaptively weight structural alignment across the hierarchy.

Crucially, because the top-level goal $\Phi^{(L)}$ is held fixed (low-entropy anchor), this entire process implements **hierarchical inverted inference**: simulating downward from goals to recover multiscale causal trajectories.
Finally, since each layer forms a cycle with its adjacent layers (via top-down structure and bottom-up reconstruction), the composition of these local cycles forms a **hierarchical chain of inference cycles**, yielding a multiscale extension of CCUP that preserves dynamic alignment between context and content at every level.

\end{proof}

\noindent\textbf{Proof of Proposition 3}
\label{appendix:H}

\begin{proof}[Sketch of Proof]
(1) Path-dependence follows from the conditional structure of $P(\Psi_t | \Psi_{t-1}, \Phi)$, which recursively updates state transitions based on prior trajectory steps. \\
(2) The manifold $\mathcal{M}_\Phi$ is constrained by both $\Phi$ and the transition model, limiting its degrees of freedom compared to the full context space. \\
(3) Under mild geometric assumptions, optimization over $\mathcal{M}_\Phi$ can be carried out with polynomial complexity in its intrinsic dimension, avoiding the combinatorial explosion of global search.
\end{proof}

%\begin{proof}[Sketch of Proof]
%Each refinement step \( \Phi^{(t+1)} := f(\Phi^{(t)}, \Psi^{(t)}) \) embeds additional structural constraints derived from past inference outcomes. As a result, the feasible set $\mathcal{M}_{\Phi^{(t)}}$ shrinks or becomes more specialized with each iteration. The chain of intersecting manifolds accumulates structure, thereby reducing entropy and increasing inference specificity. This leads to cumulative bootstrapping of representation, grounded in temporally coherent inference.
%\end{proof}

\end{document}